\documentclass[%
 ,twocolumn%
 ,secnumarabic%
,amssymb, amsmath,nobibnotes, aps, pre]{revtex4}
\usepackage[T1]{fontenc}
\usepackage[cp1250]{inputenc}
\usepackage[polish,english]{babel}
\usepackage{graphicx}
\usepackage{bm}
\begin{document}

\title{Opinion formation in networked societies with strong leaders}
\author{Paweł Sobkowicz}
\email{pawelsob@poczta.onet.pl}
\date{\today}

\begin{abstract}
Recent studies show that many types of human social activities, from scientific collaborations to sexual contacts, can be understood in terms of complex network of interactions. Such networking paradigm allows to model many aspects of social behaviour with relatively simple computer models. The present work investigates the influence of single leaders on opinion formation within  simulations of agent based artificial networked societies.  Several types of network systems (among them random networks, highly clustered, small world and scale-free) are studied. The strength of the social influence of individuals is assumed to be dictated by distance from an agent to another, as well as individual strengths of the agents. We study the effect of different topologies on the conditions of general acceptance of leader's opinion by the society.
\end{abstract}

\maketitle

\section{Modeling opinion formation in agent based computer models}

Computer simulations are quickly becoming recognized and accepted tool in social sciences, viewed as a way of testing hypotheses and predictions and revealing some simple mechanisms underlying complex behaviour. The studies are especially useful when the studied social phenomena can be quantified, and when enough empirical data exists to compare the numerical models with reality. 

One of such areas is opinion formation in societies, measurable both on the large scale (for example through election results or polls) or on small scale direct experiments. Computer simulation models of opinion forming date back to 1960s, however, the particular model we base our research on is the one of  \citet{nowak90-1}, developed further by  \citet{nowak96-1} and   \citet{kacperski99-1,kacperski00-1}, \citet{holyst01-1}. The general approach analyses the formation of public opinion through interactions between individual members of the society, taking into account differences in receptiveness, strength of influence and preexisting biases. The original work \cite{nowak90-1} has shown that interesting macroscopic behaviour can result from simple microscopical properties of computer agents. 

The basis for the models in \cite{nowak90-1,nowak96-1,kacperski99-1,kacperski00-1,holyst01-1}, which will be generally followed here, is the following:
\begin{itemize}
	\item A set of $N$ interacting agents form a society. Interactions take form of one to one activities. The dynamics of the total system is assumed to take place through discrete time steps, with updates done globally. 
	\item Each agent $i$ has, at a given time, his `opinion' $\sigma_i$. The changes of aggregates of `opinion' within the society or certain subgroups is the topic of the research.
	\item Each agent is characterized by the strength of his possible influence on other agents. This allows to model situations of uneven distribution of influence, such as presence of strong leader(s).
	\item As a special case, one of the agents (the leader) is assumed to have the strength of influence much greater than the rest of the agents.
	\item  Interaction between particular agents is governed by the strength of agents as well as the remoteness of the interacting agents. We use the term `remoteness' in abstract sense, to denote the measure of the \textit{social} separation between members of society and its effect on transmission of opinions. The overall  \textit{social impact} of agent $i$ on agent $j$  is given by combination of acting agent strength and the separation between $i$ and $j$. 
	\item The agents interact and influence each other in turns, changing their opinion after each full turn of interactions take place. Agents are allowed to interact with themselves, this mimics the phenomenon of self-support, or inclination to hold agent's present opinion.
	\item The model allows for extra-social influence or bias, applying it uniformly to all agents.
	\item The models may allow for the noise in communication and changing individual opinion by adding an equivalent of temperature to the simulations.
\end{itemize}

It is extremely interesting that the results of the works cited above have significant dependence on the way the `remoteness' between the agents is introduced. The works of Nowak et al. \cite{nowak90-1, nowak96-1} have used simple spatial two-dimensional (2D) model of agent distribution. It has been argued that 
\begin{quote}
	`People are more likely to interact with neighbors, that is those who live close to them in physical space' \dots
	`Our choice  of a 2D lattice represents quite well the physical distribution of people on flat surfaces. The results of studies conducted in Boca Raton, Warsaw and Shanghai, have shown that the probability of social interactions is decreasing as a square of physical distance' (\cite{nowak96-1}).
\end{quote}
 Results obtained for the 2D lattice based society have shown that there are localized `bubbles' of uniform opinion around strong leaders, growing and merging or diminishing with changes of external influence.
The geometrical model on a 2D disc has been used by \citet{kacperski99-1}, with the measure of decrease of the influence with distance studied mostly using linear relation.

In their later works of Kacperski, Hołyst and Schweitzer \cite{kacperski00-1,holyst01-1} have argued that the social distance, which does not have to fulfill the same conditions as geometrical one (such as, e.g. triangle inequality) should be modelled by more general and flexible model. The authors introduce the notion of social \textbf{immediacy} $m_{ij}$  (between agents $i$ and $j$ ). The immediacy $m_{ij}$ does not need to be equal $m_{ji}$ and the distribution of values of $m_{ij}$ may be arbitrary. In \cite{kacperski00-1} three immediacy distributions were analysed: uniform random distribution $0 \ge m_{ij} \ge 2\overline{m}$,  exponential distribution and discrete multimodal distribution. The results show rapid jumps in majority-minority proportion of opinion and presence of local clusters of opinion, which in abstract social distance space correspond to Nowak's bubbles of opinion.

\section{Opinion formation in networked societies}

\subsection{Network systems -- an overview}
Recent   studies of  various networked systems (see for example \citet{strogatz01-1,albert02-1, dorogovtsev02-1, dorogovtsev02-2, dorogovtsev02-3, newman00-1, newman03-1, newman03-3}), including biological and social systems, the Internet and World Wide Web have shown several universal and interesting effects. The most important among them are `small world effect', degree distributions exhibiting power-law or truncated power-law behaviour  and significant clustering effects. Large number of results indicate that the formation and properties of networked systems found in nature --- including social networks --- shows certain regularities and rules. In this paper we study the same basic model of opinion formation as Nowak, Hołyst, Kacperski and co-workers, but we place the agents on nodes of several specific networks, differing in topology of connections. The different network topologies result in significant differences in network characteristics, which might influence the process of opinion formation. 
The main networks used in this work are described below.

In random (\textbf{RAND}) networks the links are distributed randomly, thus we have a meshed network of links, with agents differing in number of the neighbours, and no general structure. On the average, the maximum distance between agents in random network grows as $\log N$ and  the number of nieghbours per agent is given by the Poisson distribution.

The nearest neighbor (\textbf{NN}) nets are formed by linking together a fixed number of closest neighbors. The easiest way to visualize such network is to place the agents on an imaginary circle and connecting each agent to $n$ neighbors.  Interesting property of NN networks is that for small filling factors, agents on the opposite points at the circle to communicate must go through many intermediaries. For 4000 agents and number of neighbors set at 10, the longest `distance' is 200 `hops'. Due to such large separation any change of behavior of the agent $i$ is seen immediately only by his closest neighbors  but the majority of the society is affected only after  filtering by intermediate agents located far from $i$.  

The small world Strogatz-Watts (\textbf{SW}) networks, introduced and popularized in recent years, reproduce a curious fact observed in many natural and human-produced networks, namely that the distance between any two nodes of the network, measured in number of links needed to connect them, is usually much smaller than that in  nearest neighbor or even random networks of the same filling ratio \cite{milgram67-1}. The name of the network category comes exactly from such observation. One of the ways to realize a SW network is through a simple reworking of the NN model: one cuts a small (even very small!) fraction of the links from between nearest neighbors and applies them instead between random agents. Keeping in mind the visualization of NN networks as nodes along a circle, this corresponds to adding connections that criss-cross the circle at random. Due to such shortcuts, even if their number is very small, the average distance between any two nodes drops dramatically. Thus we have a network that for each agent, locally is very similar to NN model (as most of the neighbors are, in fact, the same), but globally the communication through the network is much  faster.
   
The scale-free Albert-Barabási network (\textbf{AB}) reproduce another effect found in natural and artificial systems. In AB networks the dispersion of the number of immediate neighbors (degree distribution) scales according to power law rather than exponential or Gaussian. This type of networks results from interplay of two processes: growth of the network and preferential attachment (the `rich get richer' principle). As a result, the AB networks exhibit much greater presence of highly connected agents than other types of networks considered here.

We study also two types of hierarchical networks, simple one (\textbf{HS}), with each agent having one link up (to its `parent' on higher hierarchy level) and fixed number of links down (to descendants), and clustered hierarchical network (\textbf{HCL}), in which in addition to links up and down, the agents having the same parent are all linked together (within the same level). While the simple hierarchy shows no clustering at all, the clustered version preserves the division of the society into separate groups and levels, while providing high clustering ratio within groups

For completeness, in addition to the abstract space networks, we study also two traditional spatial geometries, 2D square array (\textbf{2DSQ}) and 3D cubic array (\textbf{3DCU}). In both cases agents may be linked with all immediate neighbors, including diagonal ones (thus the 2D agent can have 8 neighbours, and in 3D case up to 26 neighbours). We have used periodic boundary conditions.

In our simulations  we have tried to compare results for different networks but the same  average number of neighbors $N_N$. For the regular 2D and 3D networks, where the geometrical number of neighbours is fixed, to achieve the value of $N_N$ correspondig to other networks, we have randomly removed appropriate number of links.  The only exceptions were  the hierarchical structures, where to preserve the essential top-down asymmetry of the network, we have had to accept the lower number of connections at the lowest level  (edge of the network due to finite size), where only links up (HS) or links up and in horizontal cluster (HCL) are present. The hierarchical networks have therefore smaller average number of links, but preserve the $N_N$ within the core of the network. 

\begin{figure*}[!h]
\centering
\includegraphics[height=24cm]{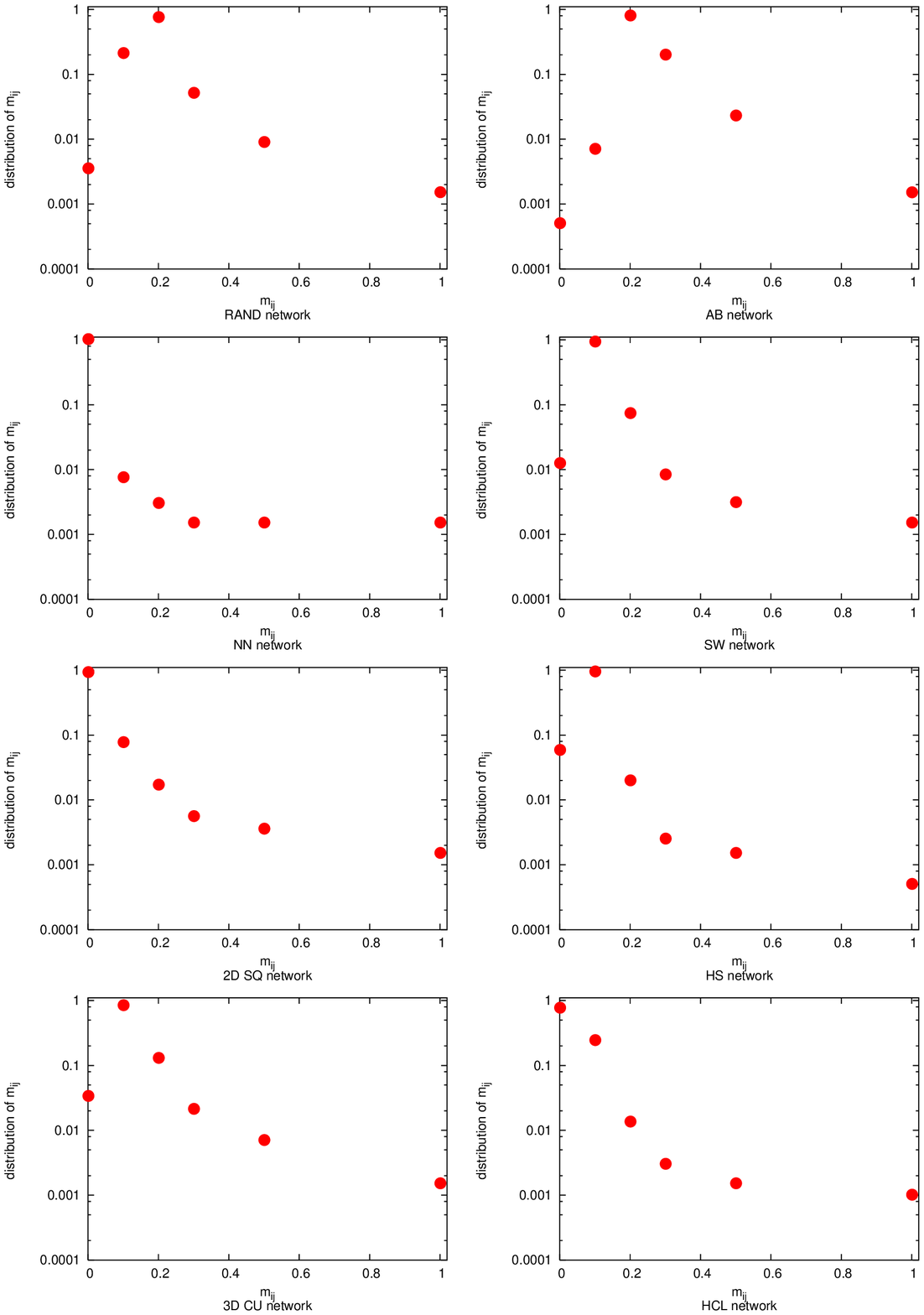}
\caption{Distribution of $m_{ij}$ for various types of networks  
\label{figure0a}}
\end{figure*}

\begin{figure*}[!h]
\centering
\includegraphics[height=24cm]{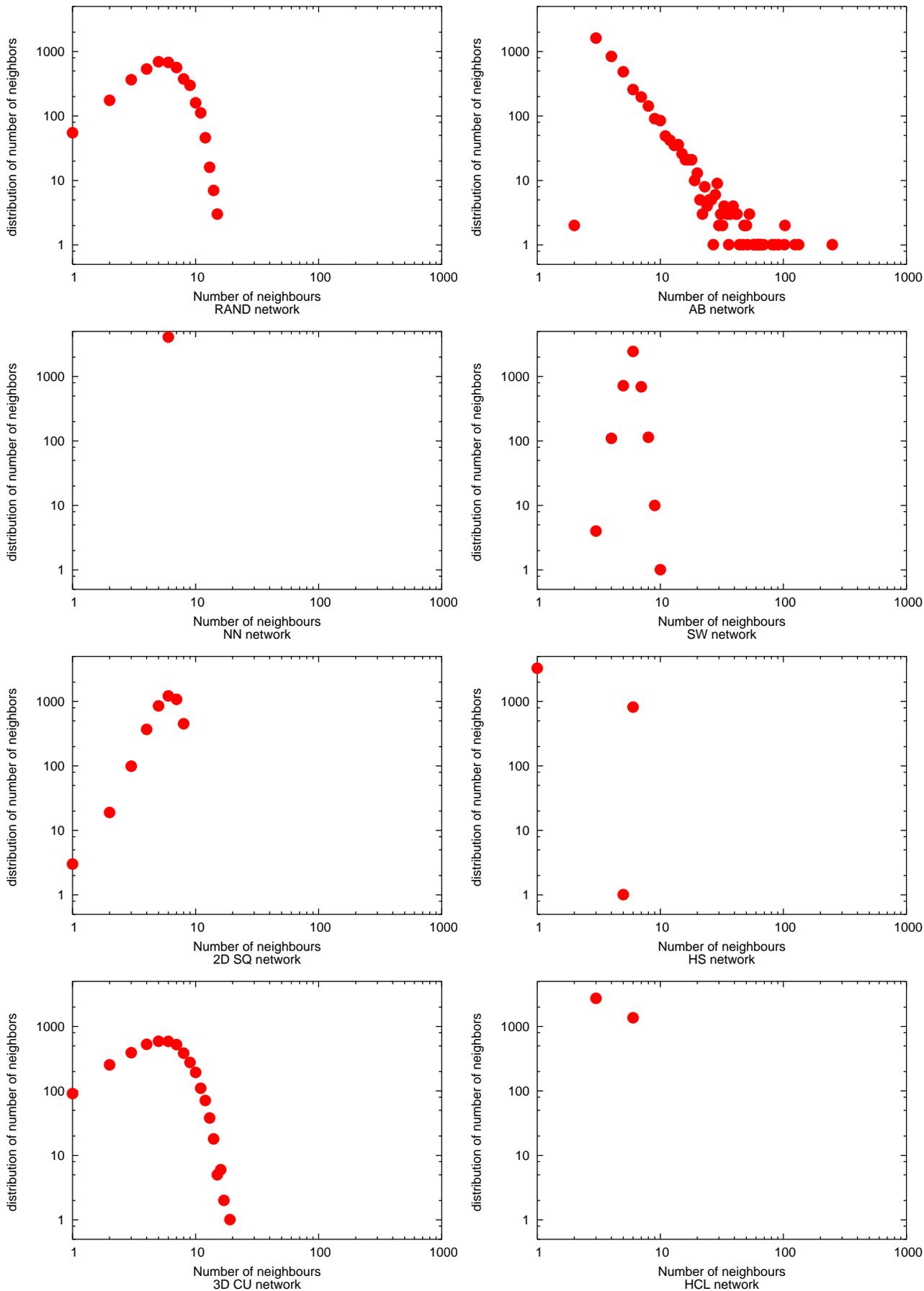}
\caption{Distribution of the number of neighbours for various types of networks  
\label{figure0b}}
\end{figure*}

Figures~\ref{figure0a} and \ref{figure0b} compare the two main properties of the discussed networks relevant to the present research. The results shown were calculated for the average number of neighbours $N_N=6$. The first figure compares the distribution of immediacies calculated according to Equation~\ref{eq:mij} ($m_{ij} = 1/ d_{ij}$, i.e. with $\alpha =1$)  for various networks. In all cases one sees very pronounced maxima in distribution of $m_{ij}$ (notice the lin-log scale), but the most probable values differ: they are the highest for the RAND and AB networks, smaller for SW, 3DCU and HS  networks, and very small for NN, 2DSQ and HCL. Increasing $\alpha$ would shift the distributions of $m_{ij}$ to even smaller values. It is worth noting that in the NN system case due to very large distances most $m_{ij}$ values are  very small indeed.

Examining the distributions of the number of neighbours one sees that most networks show some limited dispersion around the average $N_N=6$, the only exception being the scale-free network of Albert-Barab\'asi. Out of the `normal' networks, the dispersion is largest for RAND system, while for NN network the number of neighbors is fixed $N_N \equiv 6$. The AB network contains quite a number of agents with high connectivity, even up to 250 in this particular case. This turns up as the origin of one of the interesting results of our simulations.

\subsection{Use of networks in modelling opinion formation}
Even if we concentrate only on studies of strictly social phenomena, there is growing body of data that the communication processes in human societies are best modelled by complicated network arrangements (for examples see \cite{albert02-1,newman03-3}). The  `small world' theorem, well established experimentally, stating that the separation between randomly chosen members of society is much smaller than expected from geometrical and NN models, and even from randomly linked model, can in obvious way influence the process of communication between members of society, and thus the formation of opinions, or the `distance' from leaders to `normal' people. On the other hand, the presence of extremely highly connected nodes (seen in Albert-Barab{\'a}si models) gives particular influence to some individual members of the society. 

The aim of the current work is to present some results of the same type of simulations used before on geometrical or abstract connections topologies applied here on various types of networks. The differences between the general network properties might shed some light on the way the opinion shift occurs, such as the Nowak's `bubbles of opinion' in 2D model.

\section{Details of the model}
\label{sec:DetailsOfTheModel}

As mentioned above, our model followed almost exactly that of Hołyst and Kacperski. There are $N$ agents in the simulated system (we have used $N=4096$). Each agent $i$ is described by strength $s_i$ (unchanged during the simulation run) and the value of opinion, described by $\sigma_i = \pm 1$. The process of establishing the opinion in the society is modelled in discrete time steps. At each step the opinion of any given agent is updated in accordance with the combined influence of the other agents, agent's own self-influence and possible external conditions. The interaction between agents, for example agent $j$ influence on agent $i$ is given by combination of agent $j$ strength $s_j$ `remoteness' of the agents, described by immediacy $m_{ij}$. In our work we set a very natural model for $m_{ij}$ in networked system
\begin{equation}
\label{eq:mij}
	m_{ij} = \frac{1}{{d_{ij}}^{\alpha}}
\end{equation}
where $d_{ij}$ is the distance between agents $i$ and $j$ measured in number of network links and the exponent $\alpha$ determines the ratio of decrease of immediacies with increasing distance (we have simulated societies setting $\alpha = 1, 2, 3$. As $d_{ij} = 1, 2, 3 \dots$ the immediacies are always $\leq 1$. The self-influence terms $m_{jj}$ were set at 2 for all agents, which corresponds to relatively high tendency to preserve one's own opinion.

The strengths $s_i$ were chosen randomly from interval $[0,2\bar{s}]$. The only exception was a single agent, called the leader, for whom the strength $s_L$ was much greater than the average $\bar{s}$. Moreover, we have fixed $\sigma_L =1$ and set $m_{LL} \gg 1 $ so that the leader's opinion $\sigma_L$ would remain unchanged during the simulation runs.  

The states of the agents are updated synchronously in discrete time steps according to the following formula:
 \begin{equation}
	\sigma_i(t+\Delta t) = 
	\begin{cases}
		1 & \text{with probability } \frac{\exp(I_i/T)}{\exp(I_i/T)+\exp(-I_i/T)} \\
		   &  \\
		-1 & \text{with probability } \frac{\exp(-I_i/T)}{\exp(I_i/T)+\exp(-I_i/T)}  
	\end{cases}	
\end{equation}
where $T$ is the measure of randomness (`social temperature') and
\begin{equation}
	I_i = \sum_{j=1}^N s_j\, m_{ij}\, \sigma_j + h.
\end{equation}
The value of $h$ measures uniform external influence on the system.

It is useful to introduce here the rescaled values. The scaling is given by
\begin{eqnarray}
	s_{L}^R & = & \frac{s_{L}}{N \bar{s}}, \\
	h^R & = & \frac{h}{N \bar{m}\bar{s}}, \\
	T^R & = & \frac{T}{N \bar{m}\bar{s}},
\end{eqnarray}
with averages $\bar{m}, \bar{s}$ excluding the leader. Additionally, we define the reference background field $B = N \bar{s} \bar{m}$, which is the maximum value of the background influence of all non-leader agents if they all have $\sigma_j = 1$. Now the social impact on agent $i$, in terms of the rescaled values is given by
\begin{equation}
\label{II}
I_i = s_{L}^R \frac{m_{iL}}{\bar{m}} B + \sum_{j \ne L} s_{j}^R \frac{m_{ij}}{\bar{m}} \sigma_j B + h^R B.
\end{equation}

The fact, that through the Equation~\ref{eq:mij} the network itself establishes values of the immediacies $m_{ij}$ allows to connect topological properties of the networks (which in turn may be closely related to conceptual characteristics of the society) to the opinion formation process. Thus, our work is much closer to possible experimental and practical uses than the previous research.

\section{Results}
\label{sec:Results}

In our simulations we were particularly interested in searching for conditions under which the leader can convince a significant part or the whole society to his views. We have studied the fraction $f$ of leader's supporters as function of unfavorable external influence ($h^R < 0$) and  leader's strength (rescaled) $s_L^R$. We have compared results of simulations performed with three types of starting conditions: assuming that initially all agents have the same opinion as the leader ($\sigma_i \equiv 1$), random initial distribution of $\sigma_i$ and assuming that all agents disagree initially with the leader ($\sigma_i \equiv -1$). Results of the three types of initial conditions are presented as rows in figures.

\begin{figure*}[!h]
\centering
\includegraphics[height=20cm]{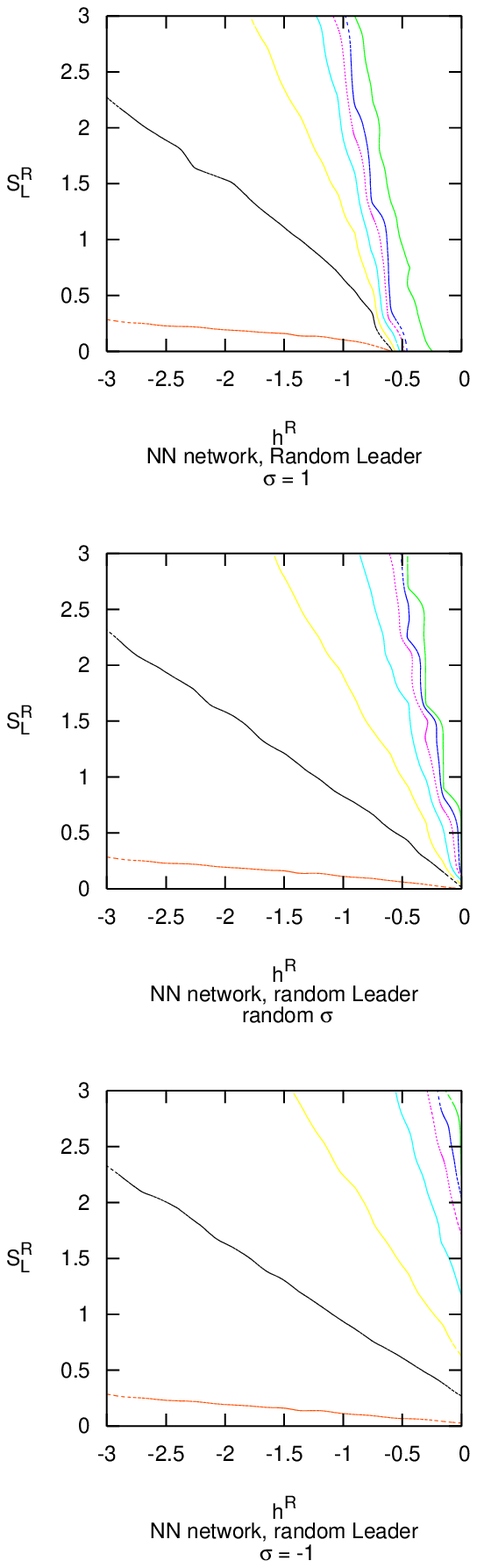}
\caption{NN network. Fraction of leader supporters $f$ compared to society size,  as function of the rescaled leader strength $s_L^R$ and external conditions $h^R$. Contours correspond to $f = 0.01, 0.10, 0.25, 0.50, 0.75, 0.90, 0.99$. $T=0.3$ and $\alpha = 1$. 
\label{figure1}}
\end{figure*}

\begin{figure*}[!h]
\centering
\includegraphics[height=20cm]{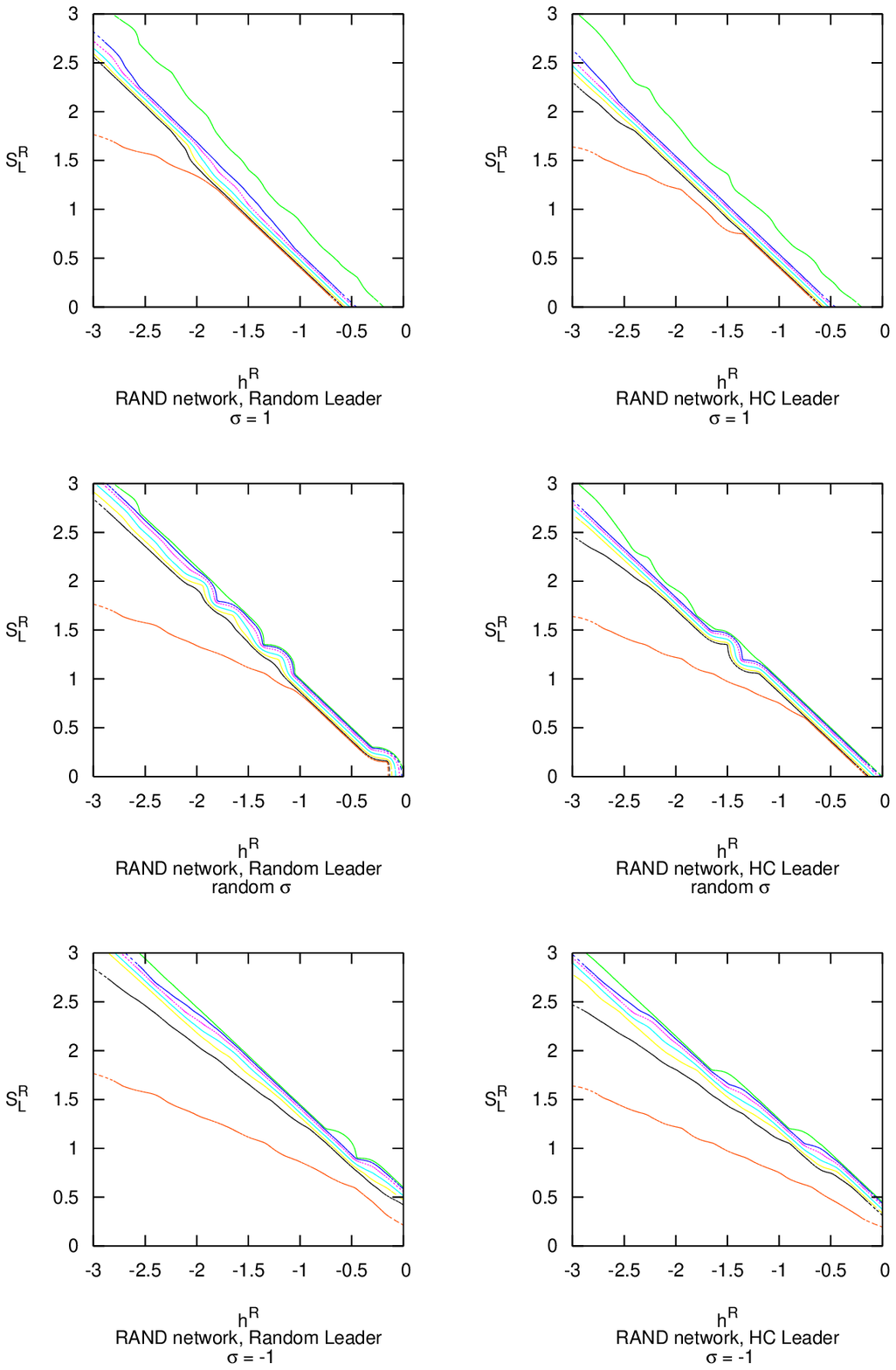}
\caption{RAND network. Fraction of leader supporters $f$ compared to society size,  as function of the rescaled leader strength $s_L^R$ and external conditions $h^R$. Contours correspond to $f = 0.01, 0.10, 0.25, 0.50, 0.75, 0.90, 0.99$.  $T=0.3$ and $\alpha = 1$. 
\label{figure2}}
\end{figure*}

\begin{figure*}[!h]
\centering
\includegraphics[height=20cm]{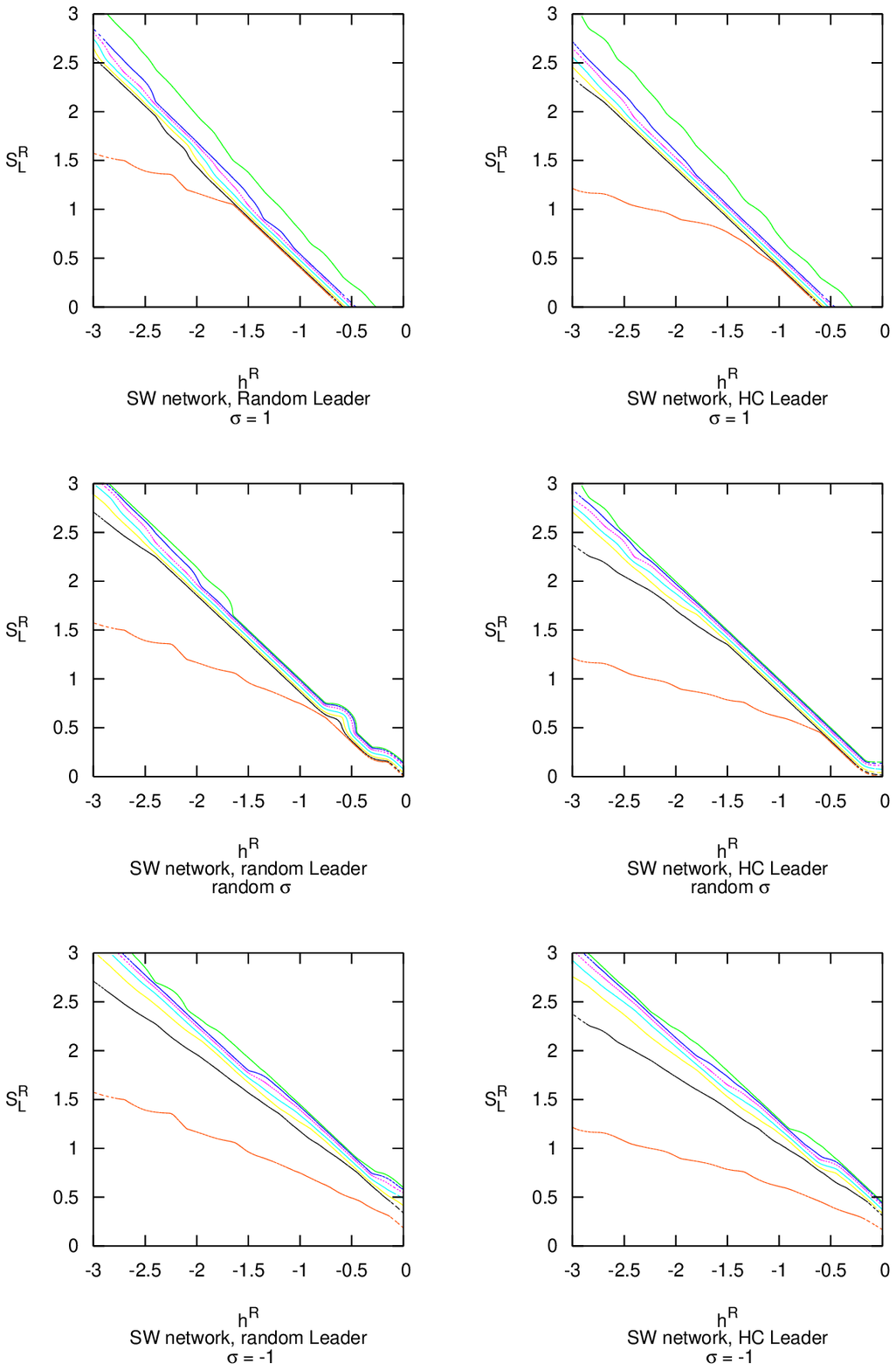}
\caption{SW network. Fraction of leader supporters $f$ compared to society size,  as function of the rescaled leader strength $s_L^R$ and external conditions $h^R$. Contours correspond to $f = 0.01, 0.10, 0.25, 0.50, 0.75, 0.90, 0.99$.  $T=0.3$ and $\alpha = 1$.  
\label{figure3}}
\end{figure*}

\begin{figure*}[!h]
\centering
\includegraphics[height=20cm]{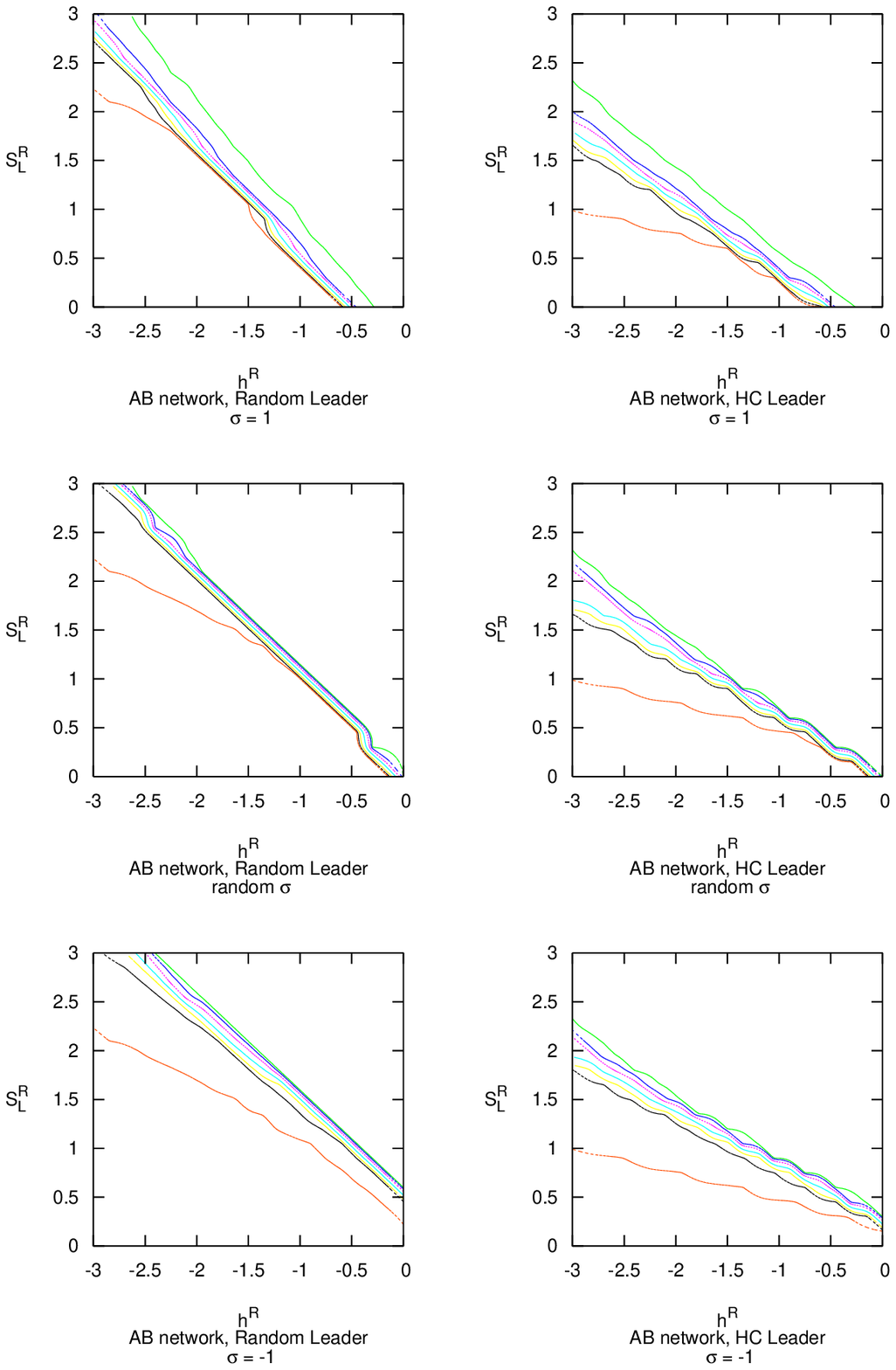}
\caption{AB network. Fraction of leader supporters $f$ compared to society size,  as function of the rescaled leader strength $s_L^R$ and external conditions $h^R$. Contours correspond to $f = 0.01, 0.10, 0.25, 0.50, 0.75, 0.90, 0.99$.  $T=0.3$ and $\alpha = 1$. 
\label{figure4}}
\end{figure*}

\begin{figure*}[!h]
\centering
\includegraphics[height=20cm]{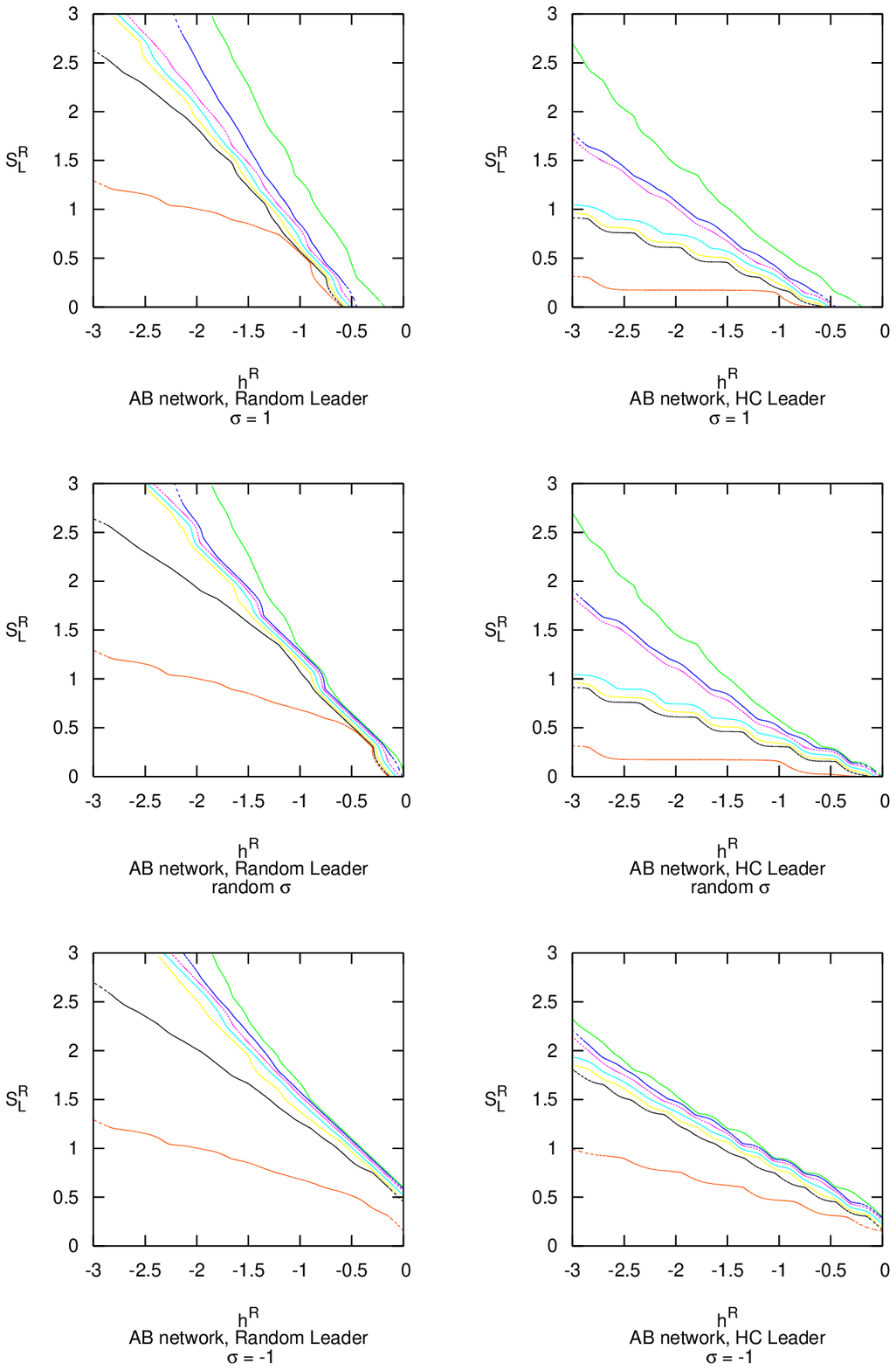}
\caption{AB network. Fraction of leader supporters $f$ compared to society size,  as function of the rescaled leader strength $s_L^R$ and external conditions $h^R$. Contours correspond to $f = 0.01, 0.10, 0.25, 0.50, 0.75, 0.90, 0.99$. $T=0.3$ and $\alpha = 2$.  
\label{figure5}}
\end{figure*}

In some of the simulations  the leader was chosen randomly, with the number of immediate neighbors close to average, in other cases  the highest connected (HC) agent was chosen to be the leader. Results are presented in the left and right columns of figures~\ref{figure1}--\ref{figure5} (the only exception being the NN network where all agents have exactly the same connectivity, so only one set of results is presented). The  number of connections of random and HC leader are rather similar in most cases --- with the exception of the scale free AB network, where there may be a difference of 1--2 orders of magnitude.

Figures~\ref{figure1}--\ref{figure5} present contour plot of fraction of leader supporters $f$ for some of the networks studied. The contours correspond to the following values of the fraction of leader's supporters: 0.01, 0.10, 0.25, 0.50, 0.75, 0.90, 0.99 (from red to green). The social temperature $T^R$ was set at 0.3. The results, are in most cases remarkably similar. The exceptions, which will be discussed below are the NN network and the AB system. 

\begin{figure*}[!h]
\centering
\includegraphics[height=20cm]{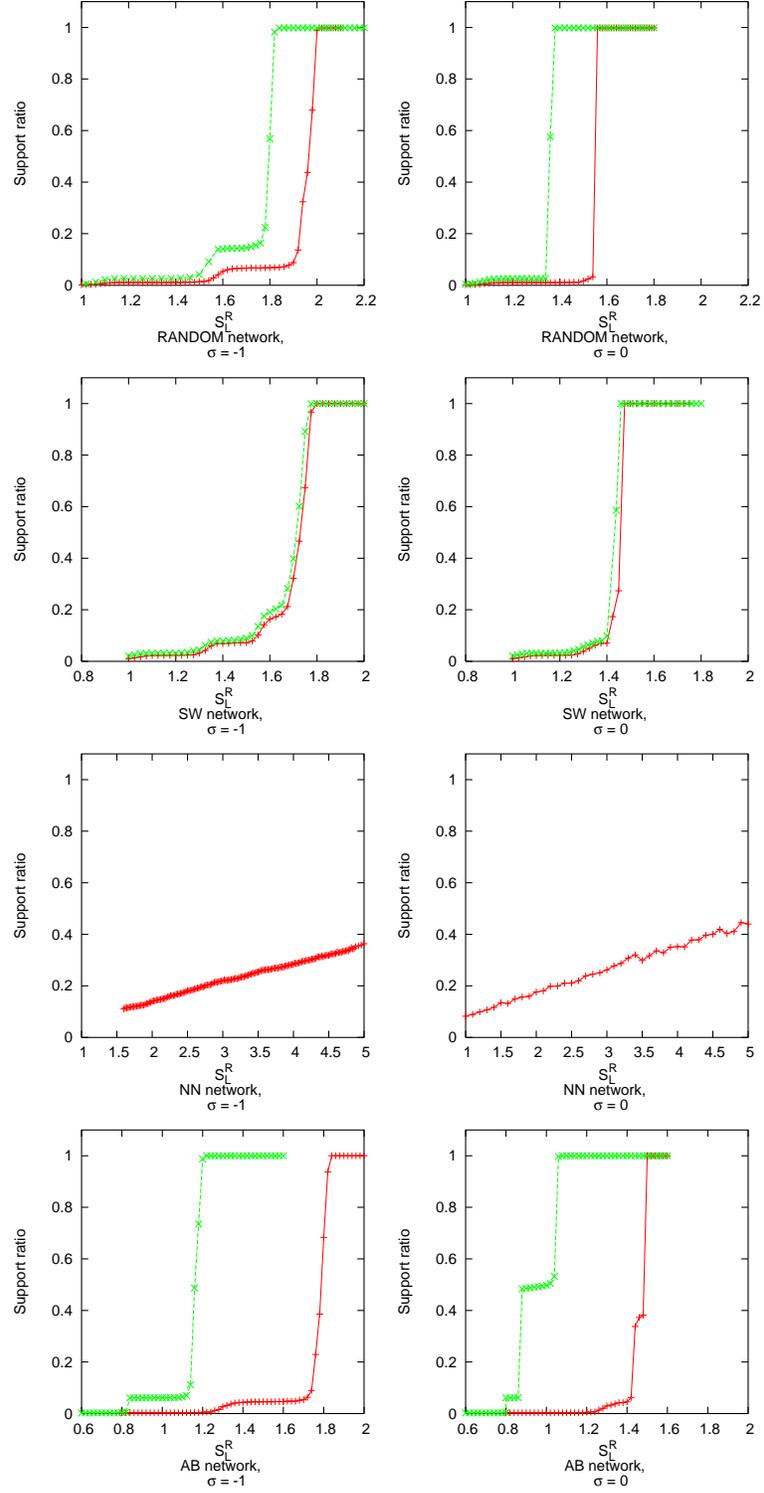}
\caption{AB network. Fraction of leader supporters $f$ compared to society size,  as function of the rescaled leader strength $s_L^R$ for fixed external conditions $h^R=-1.5$. Left column corresponds to $\sigma_j\equiv-1$ starting conditions, right column to random $\sigma_j$, with $\bar{\sigma}=0$. Red lines are the results of simulations with leader in a random position, while green lines correspond to leader as highest connectivity (HC) position.  
\label{figure8}}
\end{figure*}

Figure~\ref{figure8} presents $f$ as function of $s_L^R$ for a fixed value of $h^R=-1.5$, for a set of network types and initial conditions. Left column corresponds to extremely disfavourable initial conditions $\sigma_j\equiv-1$, right column to a random distribution with average $\bar{\sigma}=0$. Two sets of lines correspond to leaders placed at random (red) and in a HC position (green).

Let's discuss first a typical situation, such as the one found in RAND and SW networks, as well as the AB network when the leader is chosen randomly. For low $s_L^R$ and significant negative $h^R$ there are no supporters of leader's opinion. Increasing $s_L^R$ leads at first to establishment of small cluster of supporters (manifested in the area where $0.01 < f < 0.25$. In figure~\ref{figure8} this is clearly visible as plateaus of $f$. Further increase leads to rapid transition of the whole system to polarized, supporting state of $f \approx 1$. The transition region is approximately given by equation 
\begin{equation}
	s^*_L = -h^R + C,
\end{equation}
 where the value of the constant $C$ depends on the temperature and initial conditions. This equation simply reflects the strength of the leader needed to overcome the average background field of initial configuration. In low $T^R$ limit $C \approx 1$ for the $\sigma_i=-1$ starting condition and $C=-1$ for the opposite case; for random starting condition $C=0$. These results are quite straightforward: when the leader's strength overcomes the combined effect of the background influence and external conditions, the agents start to change opinion. With the increase of their number the conditions become even more favourable and transition to supportive state occurs.

That the range of values for which the local cluster of supporters is present is greater for the favourable starting conditions than in the random or disfavourable case is also quite evident: due to relatively large value of self-influence terms $m_{ii}$ in Equation~\ref{II}, it is more difficult for the leader to convert an  agent than to maintain the supporting agent's attitude.

The results are presented for moderate values of $h^R$ and $s_L^R$, comparable to 1. For much larger values the external influence and the leader's influence dominate any local statistical fluctuations due to quenched disorder. The dynamics of the system is then trivial, given, for any agent $i$ by competition between the leader term $s_L^R \frac{m_{iL}}{\bar{m}}$ and $h^R$. Due to discrete nature of distances $d_{iL}$ and therefore $m_{iL}$,  agents with the same $m_{iL}$ are convinced (for given $h^R$) at discrete values of $s_L^R$, and $f$ shows simple multi-step behaviour.

We return now to the special cases. We'll start with the NN network. Due to the presence of very large separations between agents in NN system, leader's influence on distant agents is extremely weak, and the only practical way to supportive state is the gradual `conviction' of the growing range of neighbours. As a result we see that the region of $f$ grows linearly with increasing $s_L^R$. The growth of $f$ is slow, and very large values of $s_L^R$ are needed to ensure the wholly supportive state. The NN society may be called strongly localized, and the presence of very high separation between agents clearly influences the opinion formation. It should be noted that our NN network is, in fact, a 1-dimensional (1D) network. The 2D and 3D networks, for which the average separation grows not linearly with $N$ but rather as $\sqrt{N}$ and $\sqrt[3]{N}$ respectively should weaken the effect, but still preserve it. Our simulation size of 4096 agents is, however, too small to make it visible, as the furthest distance on 2D network is 32 and for 3D case only 8. On the other hand,  for random networks, where the largest distance   grows as $\log N$ and for SW and AB networks where it is even smaller, the conditions for rapid transition are present and the such transition is observed.

The second interesting case is the AB network with the leader in high connectivity position. Such leader has a lot of neighbours with $d_{ij}=1$ and thus with $m_{ij}=1$. The influence on those agents is relatively strong, and as  these agents form a significant fraction of the society's size (in our case $250/4096 \approx 6$\%) it is much easier to achieve the supportive state
\begin{equation}
	s^*_L \ge K(-h^R + C),
\end{equation}
with $K < 1$.

Figure~\ref{figure5} presents results for the AB network with $\alpha =2$. The decrease of $m_{ij}$ with distance being much faster, there are significant differences to the $\alpha = 1$ case. The transition region is here more gradual, with visible spread of the contours especially for large negative $h^R$ region, also for the leader in random position. While the contour $f=0.90$ falls into  similar position as in the $\alpha=1$ case, the $f=0.99$ contour is shifted to much higher values of $s_L^R$ in the case of disfavourable and random initial conditions. This is due to the difficulty of converting even moderately remote agents due to rapid decrease of $m_{ij}$. 

The results presented in this work indicate that  simple models of networked societies, based on network types actually found in nature and created by humans, together with a simple formula for turning social distance to influence strength one is able to derive the picture of the conditions necessary to bring the consensus of opinion due to the influence of a single strong individual. For most of the networks the simulated behaviour is quite simple, but for some of the network topologies (especially for the scale-free Albert-Barab\'asi networks) the results are significantly different. The conditions necessary to convert the population to leader's opinion depend crucially on leader's position and connectivity in the network. Also, for strongly localized nearest neighbour networks the achievement of fully supportive state requires far greater leader strengths than for small-world or even random networks. 

As the Strogatz-Watts and Albert-Barab\'asi networks are found to be relevant in many human societies and activities (such as friendship and acquaintance networks,  sexual contacts, scientific collaborations, scientific citations, internet WEB pages links and even telephone calls structure) the  presented model may be found of importance in the analysis of the processes of opinion formation in some of these cases.

 \end{document}